\documentclass[aps,twocolumn,pra,amsmath,amssymb]{revtex4}
\usepackage{graphicx}
\usepackage{dcolumn}
\usepackage{bm}

\begin{document}
\author{Weibin Li  } 

\affiliation{Max-Planck-Institut f\"ur Physik komplexer Systeme,  
N\"othnitzer Stra$\beta$e 38, D-01187 Dresden, Germany}

\title{Directly determining  the relative phase of two coherent solitary waves in attractive Bose-Einstein condensates}
\date{\today}
\keywords{phase modulation, Bose-Einstein condensate, collapse}
\begin{abstract}
We study the dynamic structure factor of two coherent bright solitary waves in attractive Bose-Einstein condensates confined in a harmonic trap. We demonstrate that the wave function of the two solitary waves with a fixed relative phase shows interference in momentum space. The fringes are shifted depending on the values of the phase. This momentum interference can be extracted from the dynamic structure factor of the system using stimulated two-photon Bragg scattering. Thus our method provides a way to measure the relative phase directly.
\end{abstract}
 \maketitle

\section{Introduction}
Bose-Einstein condensates (BECs) with attractive two-body interaction are dynamically unstable in a trap if the number of condensed atoms reaches  a critical value \cite{baym96,victor97,bradley97,dalfovo99}. 
Generally, hefty BECs will collapse following a switch of the s-wave scattering length from a positive value (or zero) to a negative one via the use of a Feshbach resonance \cite{inouye98} . However it was found that in such case the number of remnant condensate atoms can be above the critical number \cite{strecker02,cornish06,khaykovich02,donley01}. This was explained by the generation of bright solitary waves with repulsive interactions \cite{khawaja02,salasnich02,gordon83}. These repulsive interactions have been used to indirectly infer a repulsive relative phase  $\pi/2<\theta<3\pi/2$ \cite{khawaja02,salasnich02,gordon83} between neighbor solitons \cite{cornish06,parker06}. Origin of such phases could be quantum mechanical phase fluctuations of the bosonic field operators \cite{khawaja02} due to the nonlinear modulation instability.

On the other hand, quantum and thermal excitations, which cannot be taken into account by simple mean-field theory, play a crucial role in the collapse of BECs \cite{donley01,strecker02,cornish06,khaykovich02}. These effects have been studied by effective quantum field method \cite{morgan03,milstein03,savage03,buljan05}. An interesting result arising from these studies is the absence of $\pi$-phase relation as revealed recently by \cite{wuester}. Briefly, a relative phase of $\pi$ is introduced in mean-field theory  between neighbor bright solitons \cite{khaykovich02}. Usually one expects a $\pi$ relative phase to support repulsive interaction between solitons. However, there are indications that the repulsive interaction can come from quantum (thermal) noise rather than the repulsive phase \cite{wuester}.  In addition one also doesnot get a fixed $\pi$-phase relation within mean-field theory if BECs is developed from a nonuniform initial state \cite{carr04}. It is therefore interesting to ask what a meaningful phase would entail.  A solid starting point would be mean field theory, in which $\pi$-phase is assumed \cite{salasnich02,khawaja02,gordon83}. Collisional dynamics of bright solitons in attractive BECs confined in a trap is usually based on it \cite{parker06}. Yet other phase relation is argued in \cite{carr04,wuester} as well. Consequently ascertaining the phase becomes important.

In this paper, we propose a method to measure the relative phase directly rather than to infer it indirectly from soliton interactions. In the experiment \cite{cornish06}, two bright solitary waves are separated in coordinate space. This makes it hard to measure the phase by overlapping the density \cite{andrews97}. However interference appears if the wave function is transformed into momentum space \cite{pitaevskii99}. This indicates that we can extract the relative phase from the dynamic structure factor. The dynamic structure factor provides important information about the spectrum of collective excitations and momentum distribution \cite{zambelli00}. Stimulated two-photon Bragg scattering has been used to measure the dynamic structure factor at high momentum transfer and provided high resolution and sensitivity \cite{stenger99}.  Measuring phase in momentum space is proposed by Pitaevskii et al. \cite{pitaevskii99}, who suggest to measure the coherence of two spatially separated BECs confined in a double-well potential in this way. This method can be applied for the problems we consider here as our two bright solitons are well separated in the trap \cite{cornish06}.  We calculate the dynamic structure factors for two solitons with $\pi$-phase and in-phase ($\theta=0$) using the impulse approximation \cite{hohenberg66,stamper99}. Our results show that the dynamic structure factors have distinguishable peaks corresponding to these phase relations.  

The paper is organized as follows. In sec.II, we show that the momentum density exhibits interference due to the relative phase between two bright solitary waves in a trapped BEC. The phase carried by solitary waves is successfully recovered by a maximal response of the dynamic structure factor. Only when $\theta=0$ appears a single peak, which represents a large occupation of scattered atoms with the same energy as the photon. In contrast, two peaks would be the sign of $\theta=\pi$. In sec.III, the dynamic structure factor is discussed for a case when two solitary waves move symmetrically in a trap with a small velocity. This is the general case as they are trapped in a harmonic potential and interact repulsively. Although the shapes of the dynamic structure factor are modified by the nonzero velocity, they provide distinguishable information of the concrete values of the phase also in this case. In sec.IV, we conclude the paper. 
\section{dynamic structure factor for solitary waves in an attractive BEC}
The Gross-Pitaevskii equation (GPE) is very effective in providing both qualitative and quantitative predictions for static and dynamical properties of an attractive BEC \cite{saito01,salasnich03,khawaja02}. We consider N weakly interacting bosons trapped in a quasi-one-dimensional potential $V_{trap}=m(\omega_x^2 x^2+\omega_{\rho}^2 \rho^2)/2$, where $\rho^2=y^2+z^2$ and $\omega_{\rho}\gg\omega_{x}$ are the transverse and the longitudinal trap frequencies accordingly. The dynamics of the condensate is governed by the GPE 
\begin{equation}
i\hbar\frac{\partial \psi}{\partial t}=\left[-\frac{\hbar^2}{2m}\bigtriangledown^2+V_{trap}+U_0|\psi|^2\right]\psi
\label{3dgp}
\end{equation}
here $U_0=4N\pi\hbar^2a_0/m$ is the effective coupling constant representing two-body interaction strength. $a_0$ and $m$ are s-wave scattering length and mass of atoms.  The wave function of the condensate is normalized as $\int |\psi|^2d\mathbf{r}=1$.   

The GPE is widely accepted for studying dynamics of interacting BECs at temperature $T\approx 0\text{K}$. Explicit analytical solutions are only available in a few situations. For instance, bright soliton solutions can be derived in free space using the inverse-scattering method \cite{agrawal}. Vortex solutions in rotating BECs are also available for analytical treatment \cite{vorov03,vorov05}. It is hard to obtain the exact soliton solution in the presence of harmonic trap. In this paper, we consider the soliton solution of a BEC with a Gaussian form as it is confined in a harmonic trap. The wave function of a BEC confined in an elongated axially symmetric harmonic trap is given by \cite{khawaja02,garcia98}  
\begin{equation}
\psi_0(\mathbf{r})=\sqrt{\frac{1}{\pi^{3/2}l_xl^2_\rho}}\exp\left(-\frac{x^2}{2l_x^2}-\frac{\rho^2}{2l_{\rho}^2}\right)
\label{gaussianform}
\end{equation}
where $l_i$ is the width of condensate in the $i$-direction and $a_i=\sqrt{\hbar/m\omega_i}, i=x, y, z$ are the harmonic lengths correspondingly. Note that the Gaussian wave function is close to the exact soliton solution when $l_x$ is comparable with $a_x$ \cite{khawaja02,garcia98}.
With this Gaussian ansatz, the energy of each atom is obtained readily from the GPE
\begin{eqnarray}
\label{energy}
\frac{E}{N}&=&\sum_i\left(\frac{\hbar^2}{4ml_i^2}+\frac{ml_i^2\omega_i^2}{4}\right)+\frac{U_0}{2(2\pi)^{3/2}l_xl_{\rho}^2}\nonumber\\
&=& E_0+E_{int}
\end{eqnarray}
here we have used $E_0$ and $E_{int}$ denoting the zero-point energy in a harmonic trap and two-body interaction energy. 

A homogeneous condensate with attractive interaction in free space is unstable while the zero-point kinetic energy can prevent the collapse of a trapped condensate with the finite number of particles \cite{saito01,victor97}. 
\begin{equation}
N<N_c=c\frac{\bar{a}}{|a_0|}
\label{criticalnumber} 
\end{equation}
where $c$ is a constant of order $1$ and $\bar{a}=\sqrt[3]{a_xa_ya_z}$ is the mean harmonic length.  It implies that on one hand, at given trap frequencies and s-wave scattering length, a condensate containing a finite number of particles  is stabilized by the balance of kinetic energy and negative interaction energy. On the other hand, the interaction energy is limited by a small value too. This will turn out to be helpful in calculating the dynamic structure factor. 

In the presence of two bright solitary waves, the wave function is taken as $\psi(\mathbf{r})=\psi_0(\mathbf{r-r_1})+\text{e}^{i\theta}\psi_0(\mathbf{r-r_2})$ with relative phase $\theta$. Because of repulsive interactions between two solitary waves, they will be spatially separated in a harmonic trap \cite{strecker02,cornish06}. This makes it difficult to measure the phase in coordinate space, where overlapping density profiles could induce a fringe pattern \cite{andrews97}. This obstacle can be overcome if we transform the wave function into momentum space, where two spatially separated condensates will exhibit interference \cite{pitaevskii99}. And the momentum distribution can be measured by stimulated two-photon Bragg scattering \cite{stenger99} giving the dynamic structure factor of the system. 

At large momentum transfer $\mathbf{q}$, we can use the impulse approximation to calculate the dynamic structure factor \cite{hohenberg66,stamper99}
\begin{equation}
S(\textbf{q},E)=\int d\mathbf{p}\delta\left(E-\frac{(\mathbf{p+q})^2}{2m}+\frac{p^2}{2m}\right)n(\mathbf{p})
\label{simpa}
\end{equation}
where $\mathbf{q}$ and $E$ are momentum and energy transfered from the laser pulse to the condensate and $n(\mathbf{p})$ is the momentum density of the BEC
\begin{equation}
n(\mathbf{p})=\langle\hat{\phi}(\mathbf{p})^{\dag}\hat{\phi}(\mathbf{p})\rangle
\label{smomentum}
\end{equation}
where $\phi(\textbf{p})$ is momentum distribution obtained by Fourier transformation of wave function $\psi(\mathbf{r})$
\begin{eqnarray}
\phi(\textbf{p})&=&\frac{1}{\sqrt{2\pi\hbar}}\int d\textbf{r}e^{-i(\textbf{p}\cdot\textbf{r}/ \hbar)} \psi(\textbf{r}) 
\label{singlemomentum}
\end{eqnarray}

In order to obtain the dynamic structure factor, at first we need to know the momentum density distribution. We use the Gaussian (\ref{gaussianform}) to calculate the momentum distribution. For two solitary waves moving in the $x$-direction, the wave function of the system is $\psi(\mathbf{r})=\psi_0(x+l/2,\mathbf{\rho})+\text{e}^{i\theta}\psi_0(x-l/2,\mathbf{\rho})$. $l$ is the distance between them. The corresponding wave function in momentum space is 
\begin{equation}
\phi(\textbf{p})=e^{ip_xl/2\hbar}\phi_0(\textbf{p})+e^{i(\theta-p_xl/2\hbar)}\phi_0(\textbf{p})
\label{mpsi}
\end{equation} 
where $\phi_0(\mathbf{p})$ reads
\begin{eqnarray}
\phi_0(\textbf{p})&=& \sqrt{\frac{l_xl_{\rho}^2}{\hbar^3\pi^{3/2}}}\exp\left[-\frac{l_x^2p_x^2+l_{\rho}^2p_{\rho}^2}{2\hbar^2}\right]
\label{singlemomentum}
\end{eqnarray}

\begin{figure}
\centering
\includegraphics*[width=0.9\columnwidth]{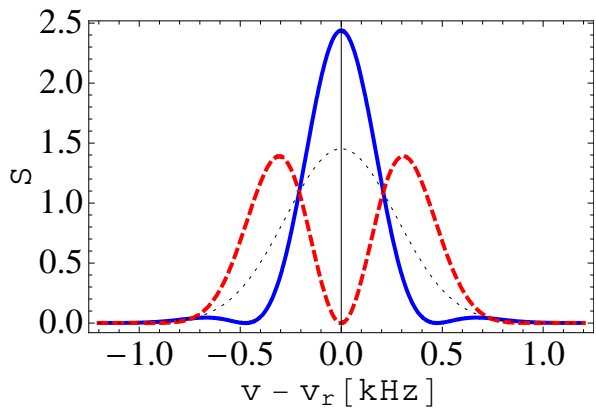}
\caption{(Color online.) Dynamic structure factor shows fringes for different phase. Solid, $\theta=0$, dashed, $\theta=\pi$; dotted, a single condensate. Other parameters are, $l=18\mu m, l_x=7\mu m, q=23\hbar\mu m^{-1}$. Here $v=E/2\pi\hbar$ and $v_r=E_r/2\pi\hbar$. }
\label{fringegaussian}
\end{figure}

\begin{figure}
\centering
\includegraphics*[width=0.9\columnwidth]{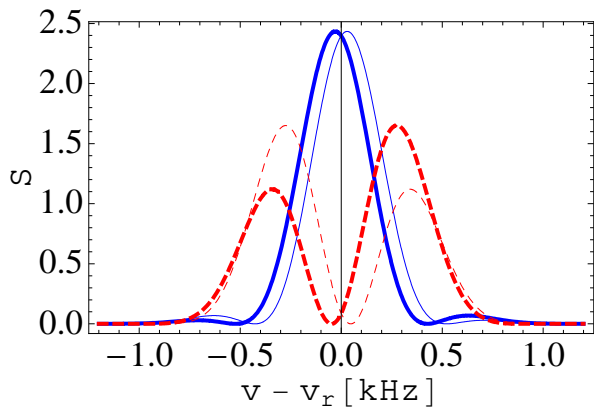}
\caption{(Color online.) Same with Fig. (\ref{fringegaussian}) but with  $\pi\times10\%$ relative phase deviation from $\theta=0,\pi$. Dynamic structure factors shift with respect to $E_r$. Thick curves show the result of $\theta+\pi\times10\%$ and the thin ones show $\theta-\pi\times10\%$. }
\label{fringeerror}
\end{figure}

For the above wave function, the momentum density exhibits interference
\begin{equation}
 n(\textbf{p})=2[1+\text{cos}(\theta-p_xl/\hbar)]n_0(\textbf{p})
\label{fringe}
\end{equation} 
where $n_0(\textbf{p})=|\phi(\mathbf{p})|^2$. The period of the interference pattern is $D_p=2\pi\hbar/l$. Obviously momentum density distribution depends on the phase $\theta$. For example, $n(p_x=0,p_y,p_z)=0$  at $\theta =p_xl/\hbar+\pi$ but has a maximum at $\theta=p_xl/\hbar$. This implies that the relative phase $\theta$ displaces the momentum interference. 

In the following, we study the dynamic structure factor of the system by applying Bragg spectroscopy \cite{stenger99}. Here interaction effects on the relative shift of the peak are negligible (Eq. (\ref{energy})). Weak interaction is particular useful because it makes  using of impulse approximation feasible at high momentum transfer \cite{zambelli00,hohenberg66}.
 
By applying a Bragg pulse in the $x$-direction,  we obtain the dynamic structure factor from Eq. (\ref{simpa})  
\begin{equation}
\label{mpal}
S(\textbf{q},E)=\frac{m}{q}\int dp_ydp_zn(p_x,p_y,p_z)
\end{equation}
with $p_x=m(E-q^2/2m)/q$. Substituting Eq. (\ref{fringe}) into the above equation, we get
\begin{equation}
S(q,E)=2[1+\text{cos}(\theta-p_xl/\hbar)]S_s(q,E)
\label{dsf1}
\end{equation}
Here $S_s(q,E)$ of a single solitary wave is
\begin{equation}
\label{sdynamic}
S_s(q,E)=\frac{ml_x}{\sqrt{\pi}q\hbar}\text{e}^{-l_x^2p_x^2/\hbar^2}
\end{equation}
Eq. (\ref{sdynamic}) has a Gaussian form. Like the momentum density, Eq. (\ref{dsf1}) also shows an interference pattern. The period of cosine function is   
\begin{equation}
\label{denergy}
D_{\text{v}}=\frac{E-q^2/2m}{2\pi\hbar}=\frac{q}{ml}
\end{equation}

At a given distance $l$ and optical momentum $q$, the relative phase $\theta$ will shift the interference pattern of the  dynamic structure factor.  Consequently, we can measure the relative phase from the fringe of the dynamic structure factor. In the experiment \cite{cornish06}, the maximal distance of two solitons is about $18\mu m$ and the width of each soliton is about $7\mu m$. Since the size $l_x$ of an attractive condensate is small, the wave length of the laser field must be far smaller than $l_x$. Here we assume the momentum of the laser field is  $q=23\hbar\mu m^{-1}$. In Fig. (\ref{fringegaussian}) we show the dynamic structure factor for two different phase relations $\theta=0,\pi$. When $\theta=0$, $S(q,E)$ is close to that of a single condensate \cite{zambelli00,stamper99}. In contrast to the maximum located at $E=E_r$ ($E_r=q^2/2m$) for $\theta=0$, $S(q,E)$ for $\theta=\pi$ it is zero there due to the interference. But two peaks emerge symmetrically around energy $E=E_r$. These results show that the dynamic structure factor exhibits contrasting interferences for $\theta=0$ and $\theta=\pi$. 

As we introduced previously, the relative phase plays a crucial role in collapsed condensates \cite{cornish06,khawaja02,parker06}. In a Bose gas with attractive two-body interaction, only a finite number of atoms can be condensed \cite{bradley97}. This is true for a single condensate. However, in a fragmented condensate, for example, several bright solitary waves interacting repulsively support a bigger system. In this case, the repulsive interaction among neighboring peaks comes from the special relative phase $\pi/2<\theta<3\pi/3$ according to mean field theory \cite{khawaja02,salasnich02,gordon83}. On the other hand, when including quantum noise, truncated Wigner quantum field simulations give $\theta=0$ rather than $\theta=\pi$ \cite{wuester}. Since the experiments did not directly monitor the phase, a measurement proposal based on the dynamic structure factor is established to distinguish the relative phase. 

We also study the dynamic structure factor if $\theta$ is slightly deviating from $\theta=0$ and $\pi$. The deviation is $10\%\times\pi$ around them.  For two bright solitons, a relative phase in the entire range $\pi/2<\theta<3\pi/2$ can stabilize the system. Such a phase deviation could be possible in the experiment. Consequently, we'd like to see how a slight phase deviation affects momentum interference and whether or not we could distinguish the phase. The results are reported in Fig. (\ref{fringeerror}).  The dynamic structure factor remains similar except a slight displacement of the peak position. For phase shifts around $\theta=0$, it only moves the maximum along the energy axis. The dynamic structure factors are not symmetric about $E-E_r=0$ any more for phase shift about $\theta=\pi$. A maximum appears in the lower energy part for a shift $\theta=(1-10\%)\times\pi$ while other smaller peaks appear in the high energy part. Vice versa for $\theta=(1+10\%)\times\pi$. Despite these shifts, the resulting momentum interference could give clearly distinguishable fringes for in-phase and $\pi$-phase.  These studies demonstrate the robustness of the dynamic structure factor against phase shift around $\theta=\pi$ and $\theta=0$. 

\begin{figure}
\centering
\includegraphics*[width=0.9\columnwidth]{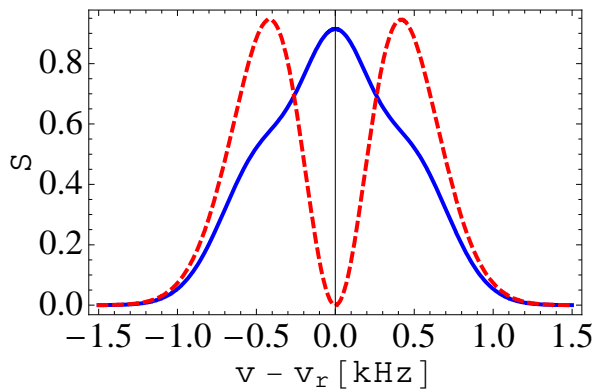}
\caption{(Color online.) Dynamic structure factor for a pair of moving solitary waves. Solid, $\theta=0$, dashed, $\theta=\pi$. Other parameters are, $l=13\mu m,v_x\approx 0.1mm/s, l_x=8\mu m, q=23\hbar\mu m^{-1}$.}
\label{velocity}
\end{figure}

\section{Dynamic structure factor for moving solitary waves}

Thus far we considered the situation when the two solitary waves have zero velocity and are well separated. However, the observed solitary waves move in the trap and the velocity cannot be always zero. Actually, the above results assume that the two solitary waves are separated by the maximal or minimal distance. In experiment \cite{cornish06}, two solitons move symmetrically about the trap center. The frequencies of the oscillation depend on the external harmonic trap frequencies. 
In this section we discuss the case when the two solitary waves have a small velocity oscillating in trap.  The wave function for a moving solitary wave is  
\begin{equation}
\psi_k(\mathbf{r})=\psi_0(\mathbf{r-r_c})e^{-ikx/\hbar}
\label{movingsoliton}
\end{equation}
where $r_c$ is the center of mass. This is a moving wave packet with momentum $-k$. Transforming it into momentum space, we get
\begin{eqnarray}
\label{velocitymomentum}
\phi_k(\textbf{p})&=& \sqrt{\frac{l_xl_{\rho}^2}{\hbar^3\pi^{3/2}}}\times\\
&&\exp\left[-\frac{2i\hbar x_c(p_x+k)+l_x^2(p_x+k)^2+l_{\rho}^2p_{\rho}^2}{2\hbar^2}\right]\nonumber
\end{eqnarray}
where $l=2x_c$. Hereafter, we assume there are two waves oscillating symmetrically in the $x$-direction with momentum $k=\pm mv$ with respect to the trap center. The dynamic structure factor for the two moving solitary waves can be obtained as
\begin{equation}
S_k(q,E)=\frac{m}{q}F_in_k(p_x)
\label{velocitysqe}
\end{equation}
where $n_k(p_x)$ is the momentum density of Eq. (\ref{velocitymomentum}) in the $x$-directions and 
\begin{eqnarray}
F_i&=&1+\exp\left(\frac{4kp_xl_x^2}{\hbar^2}\right)\nonumber \\
&&+2\cos(\theta-\frac{lp_x}{\hbar})\exp\left(\frac{2kp_xl_x^2}{\hbar^2}\right)
\end{eqnarray}
characterizes the interference of the dynamic structure factor.

\begin{figure}
\centering
\includegraphics*[width=0.9\columnwidth]{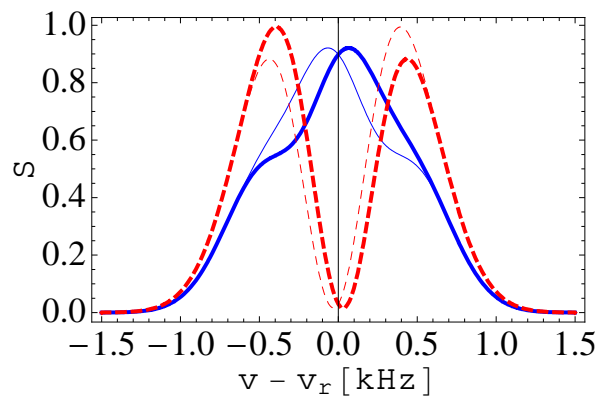}
\caption{(Color online.) Same with Fig. (\ref{velocity}) but with $\pi\times10\%$ relative phase deviation from $\theta=0,\pi$. Dynamic structure factors shift with respect to $E_r$. Thick curves show the result of $\theta+\pi\times10\%$ and the thin ones show $\theta-\pi\times10\%$. Other parameters are the same as in Fig. (\ref{velocity}).}
\label{velocityerror}
\end{figure}

For a rough estimate, the maximal velocity for a particle in a harmonic trap is $v_{max}\approx\omega_x\Delta_x$, where $\Delta_x$ is the distance to the trap center. In the experiment \cite{cornish06}, $\Delta_x\approx 9\mu m, \omega_x=2\pi\times6.8\text{Hz}$. One can get the maximal velocity if it can reach the bottom of the potential, $v_{max}\approx0.39\text{mms}^{-1}$. The velocity should be less than $v_{max}$ because of the separation of two solitary waves with repulsive interactions. Considering this point, the dynamic structure factor with nonzero velocity is plotted in Fig. (\ref{velocity}).  We can see that the interferences are different from the zero velocity cases (Fig. (\ref{fringegaussian})). The maximal momentum distribution at $E=E_r$ is lowered for $\theta=0$ and the two peaks are bigger for $\theta=\pi$ relatively. But one can still distinguish the two situations with  phase $0$ or $\pi$. 

Again, we have calculated the situation when the phase deviates from $\theta=0 (\pi)$. These results are shown in Fig. (\ref{velocityerror}), a slight shift of the maximal response will not obviously reduce  the contrast of the dynamic structure factor. Thus one can get the phase relation from measured data. 

\section{conclusion}

In this paper, we have discussed the dynamic structure factor for two bright solitary waves with different relative phases in attractive BECs. In mean field theory,  the phase $\theta=\pi$ protects the condensate from collapse by providing a repulsive force between solitary waves. However, both in-phase and $\pi$-phase relations can achieve this once quantum effects are considered \cite{khawaja02,salasnich02,gordon83,carr04,wuester}. Our study shows that momentum interference is modified by the phase. The dynamic structure factor displays distinguishable fringes at $\theta=0$ and $\pi$ respectively. At the proper photon energy and momentum, it has a single peak located at $E=E_r$ for $\theta=0$ while there are double peaks for $\theta=\pi$.  The distinguishable fringes are robust even when the solitary waves have a small velocity. $10\%$ phase deviations from $\theta=0$ and $\theta=\pi$ are also discussed. Our result offers possible way to directly measure the relative phase between two neighboring solitary waves in attractive BECs \cite{cornish06,strecker02,wuester, parker07,khawaja02}. An experiment using Bragg scattering could provide more detail information on the relative phase.

\section{Acknowledgments} 
W. Li thanks D. M. Stamper-Kurn, M. Haque, M. Hussein and Xiaotao Xie for useful discussions as well as J. Brand and S. W\"{u}ster for many important comments.

\end{document}